\documentclass[sigconf]{acmart}
\usepackage{xcolor}
\usepackage{enumitem}
\usepackage{mdframed}
\usepackage{caption}

\AtBeginDocument{%
  }

\copyrightyear{2024}
\acmYear{2024}
\setcopyright{acmlicensed}
\acmConference[KDD '24]{Proceedings of the 30th ACM SIGKDD Conference on Knowledge Discovery and Data Mining}{August 25--29, 2024}{Barcelona, Spain}
\acmBooktitle{Proceedings of the 30th ACM SIGKDD Conference on Knowledge Discovery and Data Mining (KDD '24), August 25--29, 2024, Barcelona, Spain}
\acmDOI{10.1145/3637528.3671802}
\acmISBN{979-8-4007-0490-1/24/08}

\begin{document}

\title{RecExplainer: Aligning Large Language Models for Explaining Recommendation Models}

\author{Yuxuan Lei}
\orcid{0009-0006-3235-8674}
\affiliation{%
  \institution{University of Science and Technology of China}
  \streetaddress{Fuxing Road 100}
  \city{Hefei}
  \country{China}
  \postcode{230031}
}
\email{leiyuxuan@mail.ustc.edu.cn}

\author{Jianxun Lian}
\authornote{Corresponding authors.}
\orcid{0000-0003-3108-5601}
\affiliation{%
  \institution{Microsoft Research Asia}
  \streetaddress{Danling Street 5}
  \city{Beijing}
  \country{China}
}
\email{jianxun.lian@outlook.com}

\author{Jing Yao}
\orcid{0000-0002-0527-6095}
\affiliation{%
  \institution{Microsoft Research Asia}
  \streetaddress{Danling Street 5}
  \city{Beijing}
  \country{China}
}
\email{jingyao@microsoft.com}

\author{Xu Huang}
\orcid{0000-0003-4354-334X}
\affiliation{%
  \institution{University of Science and Technology of China}
  \streetaddress{Fuxing Road 100}
  \city{Hefei}
  \country{China}
  \postcode{230031}
}
\email{xuhuangcs@mail.ustc.edu.cn}

\author{Defu Lian}
\authornotemark[1]
\orcid{0000-0002-3507-9607}
\affiliation{%
  \institution{University of Science and Technology of China}
  \streetaddress{Fuxing Road 100}
  \city{Hefei}
  \country{China}
  \postcode{230031}
}
\email{liandefu@ustc.edu.cn}

\author{Xing Xie}
\orcid{0000-0002-8608-8482}
\affiliation{%
  \institution{Microsoft Research Asia}
  \streetaddress{Danling Street 5}
  \city{Beijing}
  \country{China}
}
\email{xing.xie@microsoft.com}

\renewcommand{\shortauthors}{Yuxuan Lei et al.}

\begin{abstract}
Recommender systems are widely used in online services, with embedding-based models being particularly popular due to their expressiveness in representing complex signals. However, these models often function as a black box, making them less transparent and reliable for both users and developers. Recently, large language models (LLMs) have demonstrated remarkable intelligence in understanding, reasoning, and instruction following. This paper presents the initial exploration of using LLMs as surrogate models to explaining black-box recommender models. The primary concept involves training LLMs to comprehend and emulate the behavior of target recommender models. By leveraging LLMs' own extensive world knowledge and multi-step reasoning abilities, these aligned LLMs can serve as advanced surrogates, capable of reasoning about observations. Moreover, employing natural language as an interface allows for the creation of customizable explanations that can be adapted to individual user preferences. To facilitate an effective alignment, we introduce three methods: behavior alignment, intention alignment, and hybrid alignment. Behavior alignment operates in the language space, representing user preferences and item information as text to mimic the target model's behavior; intention alignment works in the latent space of the recommendation model, using user and item representations to understand the model's behavior; hybrid alignment combines both language and latent spaces. Comprehensive experiments conducted on three public datasets show that our approach yields promising results in understanding and mimicking target models, producing high-quality, high-fidelity, and distinct explanations. Our code is available at https://github.com/microsoft/RecAI.
\end{abstract}




\maketitle

\section{Introduction}
Recommender systems provide the appropriate information to the right individual based on comprehending users' preferences and intentions~\cite{koren2009matrix,koren2008factorization,lian2020lightrec}. These systems have become an essential component in various online services, including e-commerce, news, and television \& movies. Embedding-based recommender models, such as collaborative filtering based on latent factors~\cite{koren2009matrix,lian2014geomf} and sequential recommenders~\cite{DBLP:conf/icdm/KangM18,lei2023practical}, 
showcase their remarkable expressiveness in representing complex signals, and have thus been extensively utilized in recommender systems. However, embedding-based models typically function in a black-box manner, resulting in a lack of explainability.

Model explainability is a crucial aspect of building reliable and trustworthy recommender systems. It offers multiple advantages, including insights into the underlying logic of systems, identification of bugs, detection of biases, and providing clues for model improvement. 
One mainstream category of techniques for model explanation involves training a surrogate model to align with original black-box model~\cite{schmitz1999ann,zilke2016deepred,lakkaraju2017interpretable,ribeiro2016should}. This surrogate model must be both human-interpretable and maintain (local) fidelity to the original model. Once trained, it serves as an effective explainer for the original model. However, existing surrogate models typically employed, such as sparse linear models and decision trees, are inherently explainable but usually compromise fidelity due to their simplicity. Furthermore, the explanations generated are often limited to basic styles, such as additive weights or multiple decision rules, and lack semantic interpretation from human-readable perspective.

Recently, large language models (LLMs) have exhibited exceptional versatility and proficiency in handling complex tasks such as question answering, code comprehension, reasoning, and instruction following~\cite{kenton2019bert,radford2018improving,radford2019language,brown2020language,dong2019unified}. 
The remarkable capabilities of LLMs present new opportunities to revolutionize various research fields within machine learning, including explainable AI. With an extensive repository of world knowledge embedded in their memory and powerful multi-step reasoning skills, LLMs are renowned for generating high-quality, human-readable explanations. As a result, the traditional paradox that self-explainable surrogate models must be simple and low-complexity may no longer hold true.

In this paper, we investigate the potential of utilizing an LLM as a surrogate model for explaining recommender systems. We begin with the traditional training approach, which primarily involves aligning an LLM with a target recommendation model. The recommendation model is pre-trained and remains unaltered during this process. The LLM is then trained to emulate the recommendation model's predictive patterns—given a user's profile as input, the LLM is fine-tuned to predict the items that the recommendation model would suggest to the user. We refer to this approach as \textbf{behavior alignment}. 

However, similar to traditional surrogate model-based approach, behavior alignment merely mimics predictive observations from outside the model, attempting to deduce what is happening within the black-box. We argue that a more profound way to explain the execution logic of models involves enabling the LLM to directly comprehend the neural layers of the recommender model. Therefore, we propose an alternative approach called \textbf{intention alignment}, wherein the embeddings (i.e., activations of neural layers) of the recommender model are incorporated into the LLM's prompts to represent user and item information, and the LLM is fine-tuned to understand these embeddings. This approach can be considered as a multimodal model, with textual words and recommendation model embeddings representing two distinct data modalities. 
Take \cite{huang2023language} from the series of vision-language multimodal models~\cite{li2019visualbert,radford2021learning,DBLP:conf/cvpr/WangBDBPLAMSSW23,wang2022vlmixer} as an example.  The image pixels are transformed into embeddings by a pre-trained vision model. The language model is then trained to comprehend the contents of the original image by incorporating these embeddings into the input context. Eventually, the language model aligns itself with the vision model's space, acquiring the ability to understand and respond to questions about the image. 

Merging the advantages of both approaches, we introduce a novel method called \textbf{hybrid alignment}. This strategy effectively counteracts the hallucination issues associated with the intention alignment approach by incorporating both explicit titles and implicit embeddings in the prompt during training and inference stages. Thus, hybrid alignment facilitates a more robust and comprehensive understanding of the recommender model, ultimately enhancing the LLM to generate highly credible explanations.
To validate the effectiveness of our proposed approaches, we conduct extensive experiments on three public datasets, examining their alignment effect and explanation generation ability. Empirical evidence demonstrates that LLMs can be successfully aligned to accurately reflect and comprehend the behavior of recommender models, highlighting their potential as a new type of surrogate model for explaining complex systems. 

Our contributions can be summarized as follows:
\begin{itemize}[leftmargin=*]
  \item We propose to align LLMs for explaining recommender models, presenting a significant potential to advance explainable AI research by overcoming the traditional dilemma of requiring surrogate models to be simple for self-explainability.
  \item To enable efficient model alignment, we introduce two distinct approaches: behavior alignment and intention alignment, each providing unique benefits. Additionally, we present hybrid alignment, a method that combines the advantages of both approaches.
  \item We rigorously evaluate these alignment approaches on three publicly available datasets, demonstrating their effectiveness in both comprehension and explanation, highlighting the potential of LLMs as a new type of surrogate model for explaining recommender models.
\end{itemize} 

\section{Methodologies} 
\subsection{Problem Formulation}
In recommender systems, users are represented by their behavioral sequences:
$\mathbf{x}_u = \langle a_1, a_2, ..., a_{|\mathbf{x}_u|} \rangle$,  where  $a_*$ represents an item that user $u$ has interacted with in the past, and items are listed in chronological order. A recommender model $f()$ learns to assign a higher score to items that users favor over those they don't: $f(\mathbf{x}_u, a_i) > f(\mathbf{x}_u, a_j)$, where $a_i$ denotes a positive item and $a_j$ denotes a negative item for the user. 
To scale large systems, the two-tower model paradigm has been extensively employed in industrial recommender systems, particularly in initial steps such as item recall. In this paradigm, users and items are separately encoded into embeddings: $\mathbf{e}_u = encoder_{user}(\mathbf{x}_u)$, $\mathbf{e}_i = encoder_{item}(\mathbf{a}_i)$, and the preference score is determined by the similarity between $\mathbf{e}_u$ and $\mathbf{e}_i$. This paper focuses on the two-tower model paradigm and leaves other paradigms such as the single-tower paradigm for future work. Given a trained recommender model $f()$, our objective is to tune an LLM $g()$ to explain the decision-making process within $f()$.

In next sections, we detail our methodologies for tuning LLMs into recommendation model explainer (RecExplainer), covering three styles: behavior alignment, intention alignment, and hybrid alignment, which are denoted as RecExplainer-B, RecExplainer-I, and RecExplainer-H, respectively. 

\begin{figure*}[tb] 
 \setlength{\abovecaptionskip}{3pt}
    \centering
    \includegraphics[width=\textwidth]{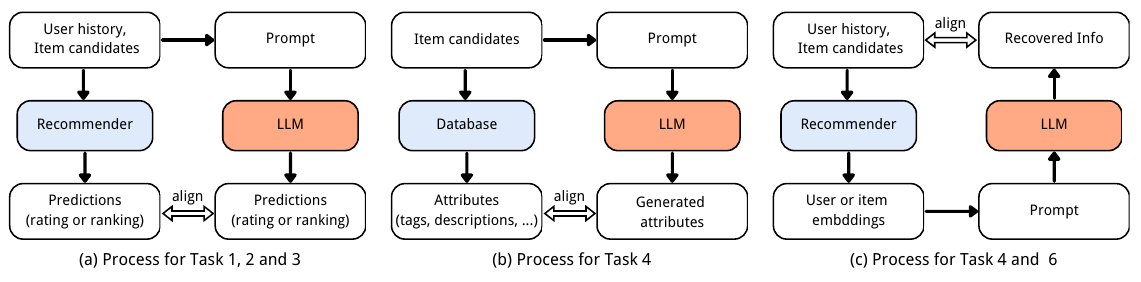}
    \caption{Graphical illustrations for aligning LLM with different tasks.} 
    \label{fig:model01}
\end{figure*}

\begin{figure}[htb]
    \centering
    \includegraphics[width=1.0\columnwidth]{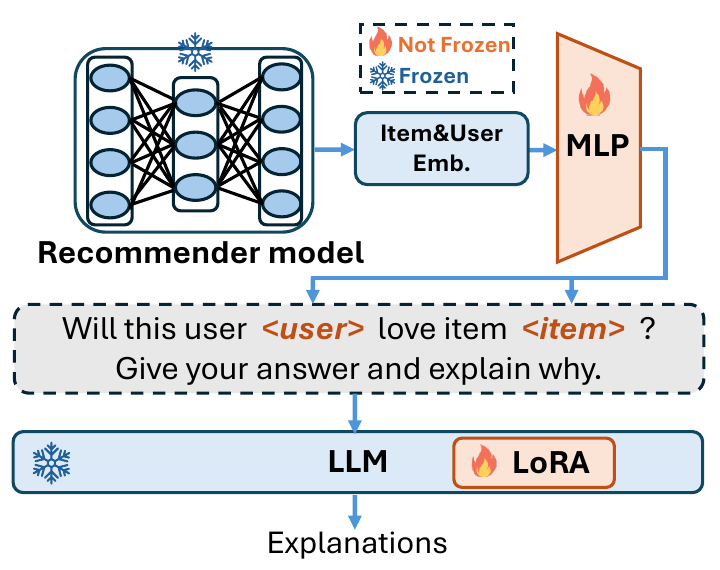}
     \vspace{-0.7cm}
    \caption{The RecExplainer framework.}
    \label{fig:model arch}
\end{figure}

\subsection{Behavior Alignment}
In this approach, we fine-tune an LLM $g()$ such that its predictive behavior aligns with the recommender model $f()$. The hypothesis is that if an LLM ideally aligns with a target model's predictions, it can imitate the execution logic of the target model and make corresponding predictions, then the LLM can leverage its inherent knowledge and reasoning capabilities to generate an explanation for its predictions. Fine-tuning tasks include:

\textbf{Task 1}: Next item retrieval. Given item titles of user history $\mathbf{x}_i$, this task teaches the LLM about the recommendations the target model would make to the user. It is important to note that there are two major differences between Task 1 for training $g$ and the traditional next item prediction task for training model $f$. First, the label in Task 1 is based on the predicted items from the target model $f$, rather than the ground truth in the original dataset. Second, the input and output content do not contain item IDs but are replaced with textual titles. These modifications ensure that the LLM focuses on understanding the target model's decision-making patterns.

\textbf{Task 2}: Item ranking. Given the item titles of user history $\mathbf{x}_i$ and a short list of item candidates $\mathbf{p}_i = \langle a_1, a_2, ..., a_k \rangle$, this task teaches the LLM to reorder $\mathbf{p}_i$ to reflect the order provided by $f()$. Retrieval and ranking are two of the most crucial tasks for recommender systems; therefore, Task 1 and Task 2 are specifically designed to align an LLM with the recommendation process of the recommender model. 

\textbf{Task 3}: Interest classification. Given the item titles of user history $\mathbf{x}_i$ and one candidate item $a_j$, the LLM must generate a binary label: like or dislike, reflecting whether user $i$ likes item $j$ or not from the perspective of $f()$. To prepare training data samples, we set up two thresholds $t^+$ and $t^-$, selecting items with $f(\mathbf{x}_i, a_j) > t^+$ as positive samples, and items with $f(\mathbf{x}_i, a_j) < t^-$ as negative samples. Task 3 serves as a complement to Task 2: while Task 2 teaches the LLM to recognize relative orders between items, it lacks the ability to discern the absolute sentiment of $f()$ towards items. 

\textbf{Task 4}: Item discrimination. A pretrained LLM may not possess sufficient knowledge about all items in a recommendation domain. This insufficiency can arise due to various reasons, such as the presence of fresh items and domain-specific items that appear less frequently in general knowledge sources. To address this issue of missing item knowledge, we design the item discrimination task: given an item title, let the LLM describe item details, including tags, descriptions, and related items\footnote{For each item, we select the top k items as its related items based on item embedding similarities generated by the target recommender model.}. This task helps the LLM to better understand the characteristics of items.

\textbf{Task 5}: ShareGPT training. To mitigate catastrophic forgetting, which leads to a decline in the LLM's general intelligence during fine-tuning, we also incorporate a general-purpose instruction tuning dataset. ShareGPT\footnote{\url{https://huggingface.co/datasets/anon8231489123/ShareGPT_Vicuna_unfiltered}} is a publicly available dataset that contains conversations between users and ChatGPT, which is gathered from ShareGPT.com with public APIs.
This dataset has been used for training numerous popular LLMs, such as Vicuna \cite{chiang2023vicuna} and MPT\footnote{\url{https://www.mosaicml.com/blog/long-context-mpt-7b-8k}}. Incorporating ShareGPT training helps preserve the LLM's general intelligence and adaptability while fine-tuning on specific tasks. This task is important because when generating explanations, not only does LLM need to understand the target recommender model, but it also needs to combine its own reasoning, instruction following and other general intelligence abilities.

All five tasks play a crucial role in generating training samples for behavior alignment. After fine-tuning, the LLM is prompted to produce model explanations. For example, given a prompt like "[some system prompts here] Given a user with history: item title$_1$, item title$_2$, ..., will you recommend item xx to the user and why?", the LLM can mimic the execution logic of the recommendation model and generate a well-informed and coherent explanation, demonstrating its understanding of the underlying recommendation process and user preferences.


\subsection{Intention Alignment}
Nonetheless, LLMs exhibit capabilities far beyond mere behavior cloning. Recently, cross-modality training has shown remarkable success in enabling LLMs to comprehend multimodal content. For example, vision-language models treat text and images as two distinct modalities. By aligning perceptions derived from text and images, the resulting LLM can effectively understand the content of images. Consequently, by leveraging the inherent reasoning abilities of the LLM, it becomes capable of providing linguistic explanations for images, such as answering the question, "\textsl{Explain why this photo is funny}". 

Building upon these insights, we treat the user and item embeddings generated from $f$ as a unique modality. This data modality captures the characteristics of items and users' preferences. Consequently, we aim to align the LLM's perceptions with those originating from the user and item embeddings. 
We term this approach "intention alignment", and its underlying hypothesis is that if an LLM can comprehend the neurons of the target model while retaining its multi-step reasoning capabilities, it holds the potential to elucidate the decision-making logic of the target model.

To establish an effective connection between the LLM and embeddings from $f$, we modify the training data for Tasks 1 through 4 by replacing item names in the query prompts with their corresponding embeddings. For instance, a prompt of Task 1 becomes "[some system prompts here] Given a user with history: [a vector of user embedding], generate the next most likely item title.".  This forces the LLM to generate accurate responses based on user and item embeddings. 
Specifically, for Tasks 1, 2, and 3, we replace the sequence of item titles in user history with a special token \textit{<user>} and map it to a single projected user embedding $\mathbf{\widetilde{e}}_u$:
\begin{equation}
\label{eq:mlp_u}
\mathbf{\widetilde{e}}_u = GELU(\mathbf{e}_u W_1 + \mathbf{b})W_2
\end{equation}
The projection operation aims to extend the original user embeddings generated by f (e.g., with a dimension of 32) to match the length of token embeddings in the LLM (e.g., with a dimension of 4096). For Tasks 2, 3, and 4, we substitute the candidate item title with a special token \textit{<item>} and map it to the projected item embedding $\mathbf{\widetilde{e}}_i$, using a projection similar to \ref{eq:mlp_u} but with a new set of parameters.

In addition to Tasks 1 through 5, we design an auxiliary task for intention alignment to enhance the information fidelity between user embeddings and the users' true history:

\textbf{Task 6}: History reconstruction. Given a projected user embedding $\mathbf{\widetilde{e}}_u$, this task recovers the titles of items in the user's history or the preference summary of the user history. We leverage GPT-4\footnote{The snapshot of GPT-4 is gpt-4-0314.\label{snapshot}} \cite{openai2023gpt4} to generate a preference summary for each user based on user history titles in advance. It is important to note that Tasks 2 and 3 primarily concentrate on understanding the relationships between user embeddings and item embeddings, but they do not sufficiently explore the self-contained information within user embeddings. Task 6 is designed to address this limitation. 

For better illustration, Figure~\ref{fig:model01} provides a comparison among the processes of different tasks to highlight the distinctions, while Figure~\ref{fig:model arch} shows the model architecture of the intention alignment.
\subsection{Hybrid Alignment}
The intention alignment approach entirely depends on user/item embeddings to decode preference-related information, which may lead to a too strict hypothesis. During the training of model $f$, a certain degree of information will inevitably be lost, such as it is hard to fully identify every item in user history from the encoded user embeddings. To mitigate the information lost, we design the third approach called "hybrid alignment", combining both the previous approaches. All Task 1 through 6 are included. for tasks that involve user history or item candidates,  hybrid alignment not only include both data forms of behavior alignment and intention alignment, but also add a new data form: simultaneously put both user history/item candidates and user/item embeddings in the query prompt. Thus, a prompt may look like: "[some system prompts here] Given a user with history: [a vector of user embedding], item title$_1$, item title$_2$, ...,  generate the next most likely item title."

\section{Experiments}

\subsection{Evaluation Strategies and Metrics}
In measuring performance of our RecExplainer on explaining recommendation models, we evaluate from two perspectives:

\subsubsection{Alignment Effect}
We first assess the LLM's alignment effect, that is, to what extent LLMs can understand neurons and predictive patterns of the target recommender model. Following the previous work \cite{DBLP:journals/corr/abs-2305-07001}, we apply the \textit{leave-one-out} strategy for evaluation. We take the last item of each user's interaction sequence as the test data and use the other records for training the LLM and target recommender model. It should be noted that when training the target recommender model, we use labels from the original dataset, but when training the LLM, we use labels inferred by the well-trained recommendation model. We evaluate four alignment tasks, including task1 (next item retrieval), task2 (item ranking), task3 (interest classification), and task6 (history reconstruction). For next item retrieval, we adopt the top-K hit ratio (HR) and top-K normalized discounted cumulative gain (NDCG) for evaluation, where we set K to 5. For item ranking, we calculate the NDCG@5. For interest classification, we use classification accuracy. For history reconstruction, we define a history coverage ratio (HCR) metric, which calculates the proportion of items in the user history that appear in the predicted sequence. During the inference of LLM, we use greedy decoding to generate texts and consider a successful output when there is a strict string match between generated item names and their ground-truth names.

\subsubsection{Explanation Generation Ability}
Considering that there is no available ground truth for the explanation, we need a new evaluation system to demonstrate the effectiveness of our method on model explainability. Specifically, we design an instruction to prompt the LLM to first evaluate the target item and then generate a coherent explanation:

\textit{"The user has the following purchase history: \{USER HISTORY\} . Will the user like the item: \{ITEM\} ? Please give your answer and explain why you make this decision from the perspective of a recommender model. Your explanation should include the following aspects: summary of patterns and traits from user purchase history, the consistency or inconsistency between user preferences and the item."} 

Following \cite{DBLP:conf/acl/WangKMLSKH23}, we implement a four-level scoring system to quantitatively evaluate the response from the LLM. Complete criteria can be found in Appendix \ref{appendix:criteria}. 

\begin{itemize}[leftmargin=*,label={$\bullet$}]
    \item RATING-\textbf{0} : Incorrect classification.
    \item RATING-\textbf{1} : Correct classification, insufficient explanation. LLM provides irrelevant explanations or provide explanations with hallucination.
    \item RATING-\textbf{2} : Correct classification, acceptable explanation with minor imperfections such as lack of persuasiveness or informativeness.
    \item RATING-\textbf{3} : Correct classification, satisfying explanation.
\end{itemize}

The evaluation criteria are formulated with a two-step approach: initially assessing correctness of the classification, followed by the assessment of explanation quality. The correctness of classification holds significant importance as it serves as an indicator of whether the LLM is formulating explanations based on an accurate understanding of the target model, rather than relying on conjecture.

Considering that human annotation is extremely time-consuming and labor-intensive, we adopt a combined approach using both human annotators and LLM annotators. Some studies \cite{DBLP:conf/acl/ChiangL23, DBLP:conf/acl/DingQLCLJB23} have already demonstrated that LLM can to some extent replace manual annotations. Since GPT-4\textsuperscript{\ref{snapshot}} is currently the most powerful LLM with strong abilities to follow instructions and perform reasoning tasks, we adopt both GPT-4 scoring and human scoring strategies to evaluate our generated explanations. More specifically, for GPT-4, we use the aforementioned evaluation criteria as prompts, inputting them into GPT-4 to generate scores. We sample 500 test cases for each dataset, calculating the mean score of each LLM on each dataset. Regarding human scoring, due to cost considerations, we select a sample of 120 test cases from a single dataset. Given that there are five different LLMs for text generation, this results in a total of 600 generated texts for human evaluation. For more details about human and GPT-4 annotations, please refer to Appendix \ref{appendix:annotation}. These human evaluation results effectively complement the GPT-4 evaluation results, providing a comprehensive assessment.

\begin{table*}[ht]
  \centering
  \caption{Performance w.r.t Alignment to the target recommender model. "N/A" represents that the method can not be applied to corresponding task.}
  \label{tab:alignment ability}
  \setlength{\tabcolsep}{4.0pt} 
  \fontsize{7.8pt}{9.9pt}\selectfont
  \begin{tabular}{@{}l|lllll|lllll|lllll@{}}
    \toprule
    \multicolumn{1}{l|}{Dataset} & \multicolumn{5}{c|}{Amazon Video Games} & \multicolumn{5}{c|}{Amazon Movies and TV} & \multicolumn{5}{c}{Steam} \\ \midrule
    \multicolumn{1}{l|}{Task} & \multicolumn{2}{c}{Task1} & \multicolumn{1}{l}{Task2} & \multicolumn{1}{l}{Task3} & \multicolumn{1}{l|}{Task6} & \multicolumn{2}{c}{Task1} & \multicolumn{1}{l}{Task2} & \multicolumn{1}{l}{Task3} & \multicolumn{1}{l|}{Task6} & \multicolumn{2}{c}{Task1} & \multicolumn{1}{l}{Task2} & \multicolumn{1}{l}{Task3} & \multicolumn{1}{l}{Task6} \\ \midrule
    \multicolumn{1}{l|}{Methods} & \multicolumn{1}{l}{H@5} & \multicolumn{1}{l}{N@5} & \multicolumn{1}{l}{N@5} & \multicolumn{1}{l}{ACC} & \multicolumn{1}{l|}{HCR} & \multicolumn{1}{l}{H@5} & \multicolumn{1}{l}{N@5} & \multicolumn{1}{l}{N@5} & \multicolumn{1}{l}{ACC} & \multicolumn{1}{l|}{HCR} & \multicolumn{1}{l}{H@5} & \multicolumn{1}{l}{N@5} & \multicolumn{1}{l}{N@5} & \multicolumn{1}{l}{ACC} & \multicolumn{1}{l}{HCR} \\ \midrule
    \multicolumn{1}{l|}{Random} & 0.0023& 0.0015& 0.6153& 0.5030 & N/A & 0.0050& 0.0030& 0.6100& 0.4987 & N/A & 0.0060& 0.0039& 0.6139& 0.4977 & N/A  \\
    \multicolumn{1}{l|}{Popularity} & 0.0077& 0.0047& 0.6683& N/A & N/A & 0.0150& 0.0088& 0.7044& N/A & N/A & 0.0321& 0.0201& 0.7971& N/A & N/A  \\
    \multicolumn{1}{l|}{Vicuna-7B} & 0.0026& 0.0014& 0.2391& 0.5026& N/A & 0.0091& 0.0062& 0.2706& 0.5011& N/A & 0.0028& 0.0015& 0.3229& 0.5000& N/A  \\ \midrule
    
    \multicolumn{1}{l|}{Vicuna-7B-ICL} & 0.0379& 0.0304& 0.2661& 0.5070& N/A & 0.0144& 0.0104& 0.3005& 0.5079& N/A & 0.0461& 0.0332& 0.2907& 0.5076& N/A  \\ 
    \multicolumn{1}{l|}{GPT4-ICL} & 0.1105& 0.0864& 0.6492& 0.6338& N/A & 0.0886& 0.0692& 0.5954& 0.5964& N/A & 0.3008& 0.2525& 0.6750& 0.5886& N/A  \\ 
    \multicolumn{1}{l|}{SASRec} & 0.6736& 0.5234& \textbf{0.8759}& 0.7768& N/A & 0.6217& 0.5025& \underline{0.8252}& 0.6541& N/A & \textbf{0.9751}& \textbf{0.8780}& \textbf{0.9577}& 0.8914& N/A  \\ \midrule

    \multicolumn{1}{l|}{RecExplainer-B} & 0.7460& 0.6260& 0.7521& 0.8365& N/A & 0.8106& 0.7027& 0.7033& 0.7818& N/A & 0.9310& 0.8100& 0.8699& 0.9554& N/A  \\
    \multicolumn{1}{l|}{RecExplainer-I} & \textbf{0.8436}& \textbf{0.6994}& 0.8299& \textbf{0.9385}& \underline{0.1162}& \textbf{0.9039}& \underline{0.7709}& \textbf{0.8290}& \textbf{0.8396}& \underline{0.1201}& \underline{0.9615}& 0.8122& \underline{0.9083}& \textbf{0.9904}& \underline{0.0659} \\
    \multicolumn{1}{l|}{RecExplainer-H} & \underline{0.8057}& \underline{0.6922}& \underline{0.8458}& \underline{0.9189}& \textbf{0.1325}& \underline{0.8773}& \textbf{0.7750}& 0.7638& \underline{0.8109}& \textbf{0.1461}& 0.9358& \underline{0.8242}& 0.9036& \underline{0.9815}& \textbf{0.0707} \\
  \bottomrule
\end{tabular}
\end{table*}

\subsection{Experimental Setup}
\subsubsection{Datasets}

We evaluate our model on three public datasets: Video Games and Movies \& TV dataset released in Amazon platform\footnote{\url{https://cseweb.ucsd.edu/~jmcauley/datasets/amazon_v2/}}\cite{ni2019justifying}, and Steam\footnote{\url{https://github.com/kang205/SASRec}}\cite{DBLP:conf/icdm/KangM18,pathak2017generating}. For the specific tasks, we generate data as follows: for next item retrieval, we treat the top-1 prediction of the target recommender model as the ground truth; for item ranking, we sample five items from the entire item set for each sample, and use the ranking order produced by the target model as the ground truth; for interest classification, we set the $t+$ and $t-$ threshold as the top 20\%, bottom 50\% respectively, and sample one positive and one negative item for each user for training and testing. The dataset details can be found in Appendix \ref{appendix:data details}.

\subsubsection{Implementation details}
Our backbone LLM is vicuna-v1.3-7b \cite{chiang2023vicuna} with a maximum context length of 1024. We employ LoRA \cite{DBLP:conf/iclr/HuSWALWWC22} for parameter-efficient tuning and leverage DeepSpeed's ZeRO-2 \cite{DBLP:conf/sc/RajbhandariRRH20} to further reduce gpu memory consumption. We use 8 NVIDIA V100-32GB GPUs and fp16 for training 10 epochs,  with a total batch size of 64. The peak learning rate is 1e-4 with a linear warmup for the first 10\% steps. We train our model using the standard language modeling objective and only compute loss on the response tokens. For the target recommender model, we adopt the powerful transformer-based model SASRec \cite{DBLP:conf/icdm/KangM18}, which is lightweight and effective. Hyperparameter tuning is performed to train SASRec on three datasets, obtaining a well-trained sequential recommender. Specifically, the embedding size is set to 256, 192, 256 on Video Games, Movies \& TV and Steam datasets, respectively, and the max sequence length is set to 9, 13, 16 respectively. During LLM's training, the target SASRec is fixed, which is used only for inferring user embeddings and item embeddings.

\subsubsection{Baselines}
For evaluating the alignment effect, we employ two statistical models, a raw LLM and three aligned models.
\begin{itemize}[leftmargin=*,label={$\bullet$}]
    \item \textbf{Random}: Sample k items uniformly from the item set for retrieval. Random shuffle the candidate items for ranking.
    \item \textbf{Popularity}: Sample k items based on the item popularity distribution for retrieval. Sort the candidate items according to the item popularity.
    \item \textbf{Vicuna-v1.3-7B \cite{chiang2023vicuna}}: An open-source LLM obtained by fine-tuning the LLaMa model on ShareGPT data. This model serves as the base model which is not fine-tuned on our in-domain dataset.
    \item \textbf{Vicuna-v1.3-7B-ICL \cite{chiang2023vicuna}}: In-context learning is an effective approach to align LLMs to do specific tasks. Specifically, when performing a specific task, we randomly select two instances from the training set of that task and place them at the beginning of the LLM's context.
    \item \textbf{GPT4-ICL \cite{openai2023gpt4}}: Same in-context learning strategy but with the currently most powerful closed-source LLM from OpenAI.
    \item \textbf{SASRec \cite{DBLP:conf/icdm/KangM18}}: a mainstream traditional sequential recommender model. To enable the SASRec to align to the target recommender model and complete several alignment tasks, we use knowledge distillation outlined in EMKD\cite{DBLP:conf/sigir/DuYZZ00LS23}. Specifically, we minimize the Kullback-Leibler divergence between the teacher logits and the student logits, where the logits represent the scores given by users for the entire item set.
    
\end{itemize}
For evaluating the explanation generation ability, we use \textbf{Vicuna-v1.3-7B} and \textbf{ChatGPT}\footnote{https://chat.openai.com/} (gpt-3.5-turbo-0301), as other aforementioned methods either are not text generation models or lack readily available examples for in-context learning. Additionally, the three proposed alignment methods themselves are suitable for mutual comparisons on both evaluation settings.

\subsection{Performance w.r.t. Alignment}
To investigate the alignment effect of LLM after training, we evaluate model's performance on four recommendation-related tasks, with the results presented in Table ~\ref{tab:alignment ability}. It should be noted that for Tasks 1, 2 and 3 in Table ~\ref{tab:alignment ability}, we use the inference results of the target recommender model as ground truth labels and the SASRec baseline in Table ~\ref{tab:alignment ability} is actually another model aligned with the target recommender model. We have the following observations:

RecExplainer-H can achieve comparable performance with the powerful SASRec, and often performs better in retrieval (task1) and classification (task3) tasks. This demonstrates that our RecExplainer's alignment training is sufficiently effective, as only thorough alignment can ensure the reliability of subsequent explanation generation. For the vicuna-7b model without alignment, the performance is unsatisfactory across all tasks. This suggests that there is still a significant gap between the target model and the LLMs, demonstrating the necessity of alignment training. In comparison to vicuna-7B, vicuna-7B-ICL with the adoption of the in-context learning method shows some improvements, but its performance remains relatively low. On the other hand, gpt4-ICL demonstrates significantly higher performance than vicuna-7B-ICL, showcasing its strong general intelligence. However, the performance of gpt4-ICL still lags far behind RecExplainer-H, as alignment training enables the LLM to thoroughly learn the recommendation paradigm of the target model on the entire dataset.

Regarding our three alignment methods, RecExplainer-B performs the worst across all tasks and datasets, suggesting that merely imitating the recommendation behavior of the target model is not an optimal solution for understanding the target model. Given that the neurons of the target recommender model (such as user and item embeddings) can inherently reflect the recommendation paradigms and collaborative signals in the target model, we can see that the performance of RecExplainer-I improves significantly when these embeddings are used as part of prompts for the LLM. For RecExplainer-H, its performance is superior to all other approaches except on next item retrieval (task1) and interest classification (task3)  tasks, which are slightly lower than that of RecExplainer-I in some datasets. A possible reason is that when both neuron signals and textual signals are added to LLM's inputs, the LLM may overly rely on the text and, to some extent, neglect the role of neurons. Overall, The performance of RecExplainer-H is very powerful, indicating that textual and neuron signals can complement each other, jointly enhancing the LLM's understanding of the target model. 


In conclusion, compared to models without alignment, LLMs with alignment training significantly enhance their predictive ability for the pre-trained target model. They can achieve comparable performance with existing alignment-based recommendation models, indicating that LLMs with alignment training have effectively learned the paradigm and neurons of the target model, making it suitable for subsequent recommendation explanation tasks.

\subsection{Performance w.r.t. Explanation}
\subsubsection{Overall Ratings}
\begin{figure}[thb]
    \centering
    \includegraphics[width=0.99\columnwidth]{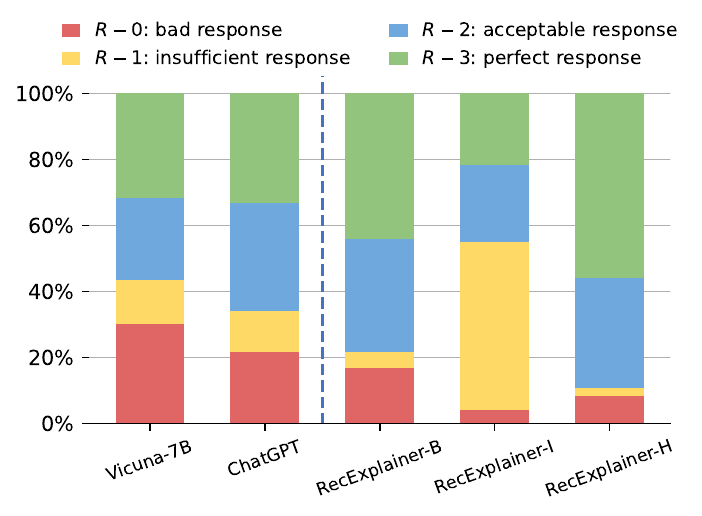}
    \vspace{-0.5cm}
    \caption{Performance w.r.t Explanation (Human experts) on Amazon Video Games dataset.}
    \label{fig:human labels}
\end{figure}

Evaluation results from GPT-4 and human experts are shown in Table \ref{tab:explan gen ability} and Figure \ref{fig:human labels} respectively. We have the following observations: 

The trend of the two evaluation strategies are the same, validating the credibility of our evaluation method. Among which, RecExplainer-H achieves the highest scores on all three datasets, indicating that it can mimic the execution logic of the target model well and give satisfying explanations for that logic. RecExplainer-B comes next, suggesting that behavioral imitation is also helpful in understanding the execution paradigm of target models.

For the two unaligned LLMs, Vicuna-7B and ChatGPT, they can generate reasonably good explanatory texts in some cases through their powerful reasoning capabilities and internal knowledge. However, since they are unrelated to the target model and are not aware of the target models' predictive patterns, they are not sure whether the target model would recommend the item or not. As a result, their explanations tend to be ambiguous and lack persuasiveness and their scores tend to fall under the RATING-2.

Another point worth mentioning is that we find RecExplainer-I has the lowest evaluation scores. By examining specific examples, we discover that it generates explanations with hallucination, such as mistaking other items as the current user's history. This indicates that there might be a certain gap in directly reconstructing textual content from neuron signals, as the information may be insufficient. This is also demonstrated by the relatively low metrics of the history reconstruction task in the previous section.

\subsubsection{Distinction and Coherence}
To further verify whether RecExplainer are indeed explaining its own predictions, we conduct validation from two perspectives: (1) Is RecExplainer's explanations distinct from other LLM's explanations? (2) Do RecExplainer's explanations reflect RecExplainer's predictions? 

\begin{figure}[t]
    \centering
    \includegraphics[width=1.0\columnwidth]{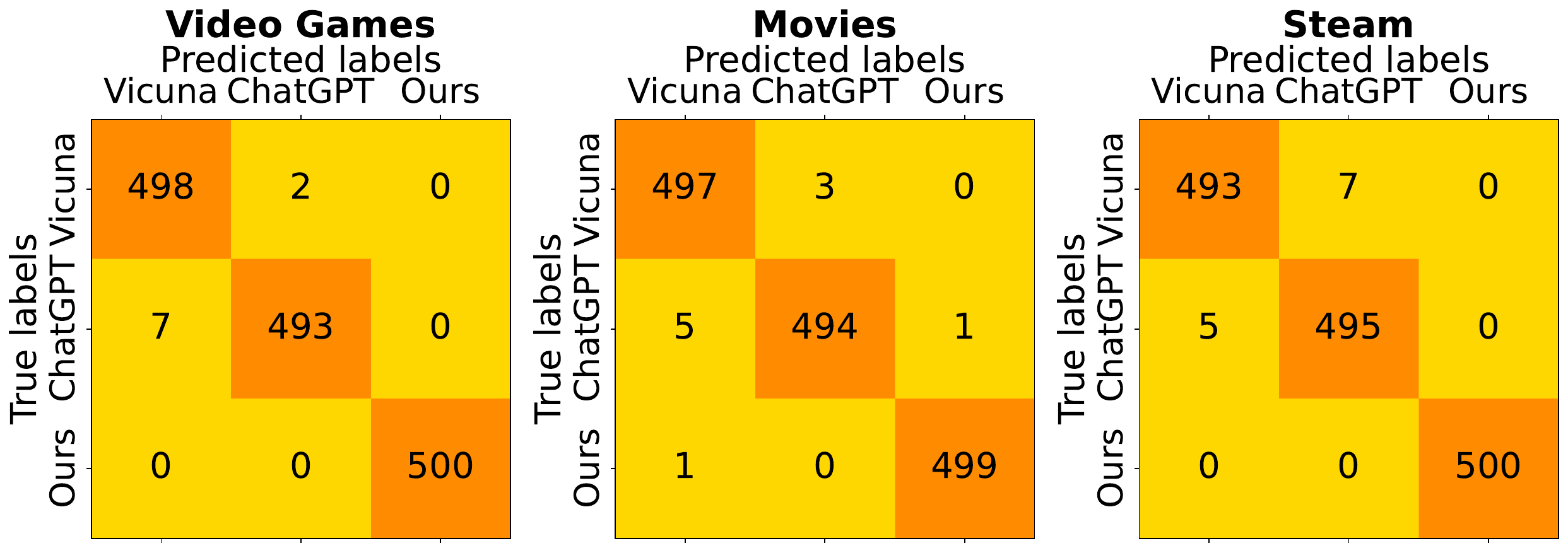} 
    \vspace{-0.5cm}
    \caption{Confusion matrix of explanation discrimination.}
    \label{fig:confusion}
\end{figure}

We generate 2500 explanations for each of Vicuna-7B, ChatGPT, and RecExplainer-H, and divide them into training and testing sets at 4:1 ratio.
Firstly, we train a discriminator to prove that the explanations generated by RecExplainer possess sufficient distinctiveness and are different from those produced by models without alignment training. The experimental results are illustrated in Figure \ref{fig:confusion}, revealing that this discriminator can easily differentiate explanations from RecExplainer and other models.
Secondly, we develop score predictors to assess the alignment between the explainer's textual explanations and the target recommender model's predictions. Specifically, for a given user-item pair $(u, i)$, $f(\mathbf{x}_u, a_i)$ represents the prediction made by the target recommender model. The score predictors are then trained using the explainer's textual explanations as input, with the goal of closely approximating $f(\mathbf{x}_u, a_i)$. This is a regression task, so we evaluate this using Mean Squared Error (MSE). The results, as presented in Table \ref{tab:ctr predictors}, indicate that explanations produced by RecExplainer offer a significant advantage when used to predict scores of the target model, confirming that RecExplainer effectively utilizes the understanding of the target model's behavior patterns during the explanation generation process. Both the discriminator and the score predictor employed in this study are based on the base version of BERT\cite{kenton2019bert}.
\begin{table}[thb]
  \centering
  \caption{Performance w.r.t. Explanation (GPT-4). The higher score represents the better performance in explanation, ranging from 0 to 3.}
  \label{tab:explan gen ability}
  \begin{tabular}{@{}l|lll@{}}
     \toprule
     \multicolumn{1}{l|}{Methods}& \multicolumn{1}{c|}{Games} & \multicolumn{1}{c|}{Movies} & \multicolumn{1}{c}{Steam} \\ \midrule
    \multicolumn{1}{l|}{Vicuna-7B} & \multicolumn{1}{c|}{2.0703}& \multicolumn{1}{c|}{2.0261}& \multicolumn{1}{c}{2.0341} \\
    \multicolumn{1}{l|}{ChatGPT} & \multicolumn{1}{c|}{1.9320}& \multicolumn{1}{c|}{1.8360}& \multicolumn{1}{c}{1.9560}\\ \midrule
    \multicolumn{1}{l|}{RecExplainer-B} & \multicolumn{1}{c|}{\underline{2.3240}}& \multicolumn{1}{c|}{\underline{2.1360}}& \multicolumn{1}{c}{\underline{2.4660}}\\
    \multicolumn{1}{l|}{RecExplainer-I} & \multicolumn{1}{c|}{1.6653}& \multicolumn{1}{c|}{1.4689}& \multicolumn{1}{c}{1.3394}\\
    \multicolumn{1}{l|}{RecExplainer-H} & \multicolumn{1}{c|}{\textbf{2.5240}}& \multicolumn{1}{c|}{\textbf{2.2204}}& \multicolumn{1}{c}{\textbf{2.4920}}\\
  \bottomrule
\end{tabular}
\end{table}

\begin{table}[thb]
  \centering
  \caption{Performance w.r.t score predictors. The metric is Mean Squared Error (MSE).}
  \label{tab:ctr predictors}
  \begin{tabular}{@{}l|lll@{}}
     \toprule
     \multicolumn{1}{l|}{Methods}& \multicolumn{1}{c|}{Games} & \multicolumn{1}{c|}{Movies} & \multicolumn{1}{c}{Steam} \\ \midrule
    \multicolumn{1}{l|}{Vicuna-7B} & \multicolumn{1}{c|}{3.6970}& \multicolumn{1}{c|}{2.8373}& \multicolumn{1}{c}{0.9687} \\
    \midrule
    \multicolumn{1}{l|}{ChatGPT} & \multicolumn{1}{c|}{3.4803}& \multicolumn{1}{c|}{2.8393}& \multicolumn{1}{c}{1.0288}\\ \midrule
    \multicolumn{1}{l|}{RecExplainer-H} & \multicolumn{1}{c|}{\textbf{1.2786}}& \multicolumn{1}{c|}{\textbf{1.9190}}& \multicolumn{1}{c}{\textbf{0.3248}}\\
  \bottomrule
\end{tabular}
\end{table}

\subsection{Case Study}
\subsubsection{Explanation Quality}
\begin{figure}[htb]
    \centering
    \includegraphics[width=1.05\columnwidth]{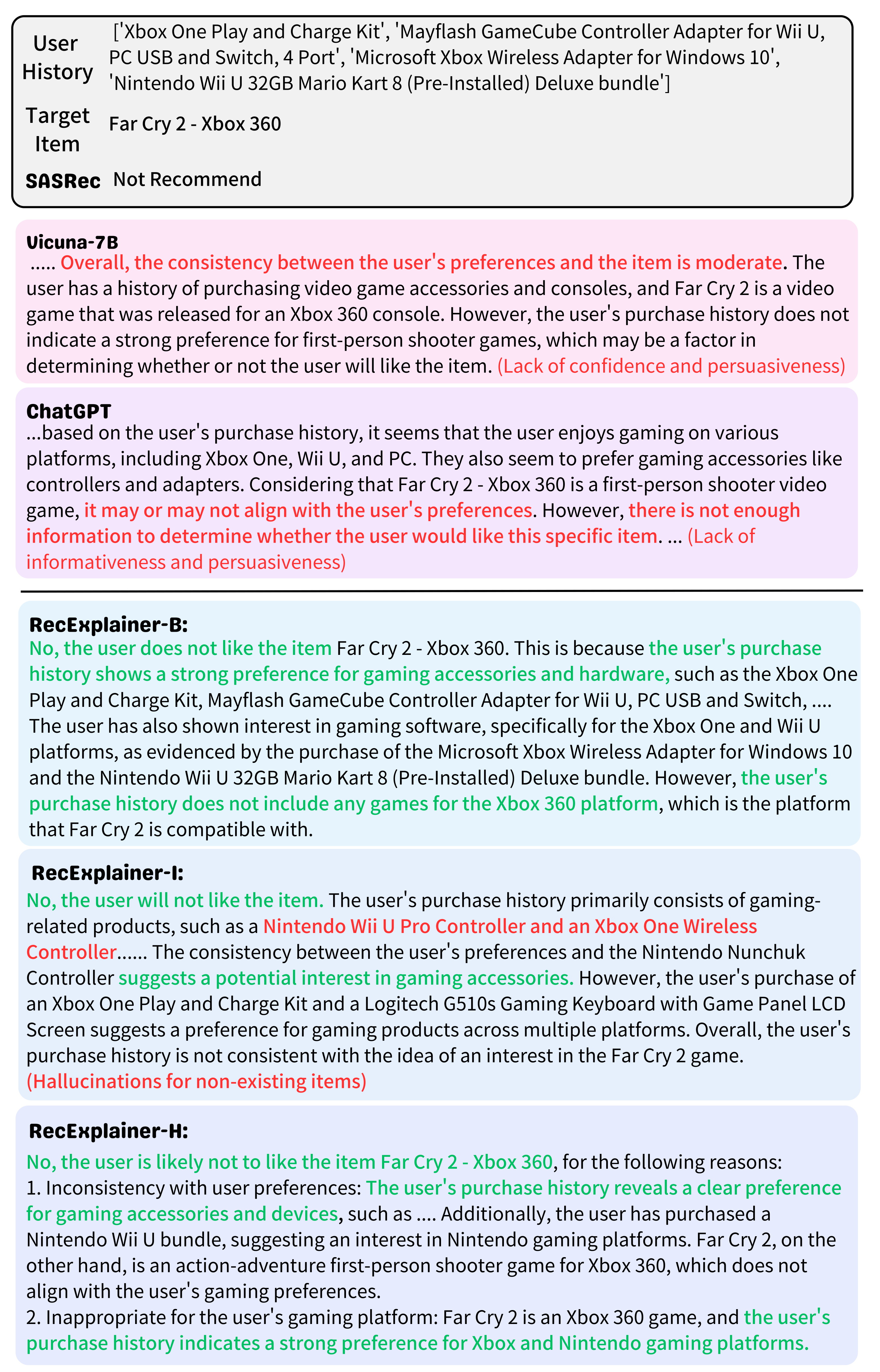}
     \vspace{-0.4cm}
    \caption{Case study for explanation on Amazon Video Games dataset. Fonts highlighted in red signify the presence of errors within the generated explanations, while those highlighted in green denote accurate and logical explanations.}
    \label{fig:case_explain}
\end{figure}

We show cases of each method for straightforward effect of explanation, illustrated in Figure~\ref{fig:case_explain}. Both RecExplainer-B and RecExplainer-H give convincing explanations, which point out that firstly the user prefers gaming accessories and devices instead of games, and secondly the user has no explicit engagement in Xbox-360 platform, exhibiting high consistency with the output of the target recommender model. Nevertheless, Vicuna and ChatGPT do not give a satisfying and persuasive explanation. This is because they do not align with the target model and can only rely on their own knowledge and logic to make certain conjectures about user preferences, which may cause errors. Notably, RecExplainer-I exhibits hallucination in giving non-existing games. This illustrates that although the alignment is effective, solely relying on hidden neurons to recover the domain information of user history/items and make explanations are not enough due to the information compression loss in embeddings of the target model.

\subsubsection{Controllable Explanations}
\begin{figure}[htb]
    \centering
    \includegraphics[width=1.05\columnwidth]{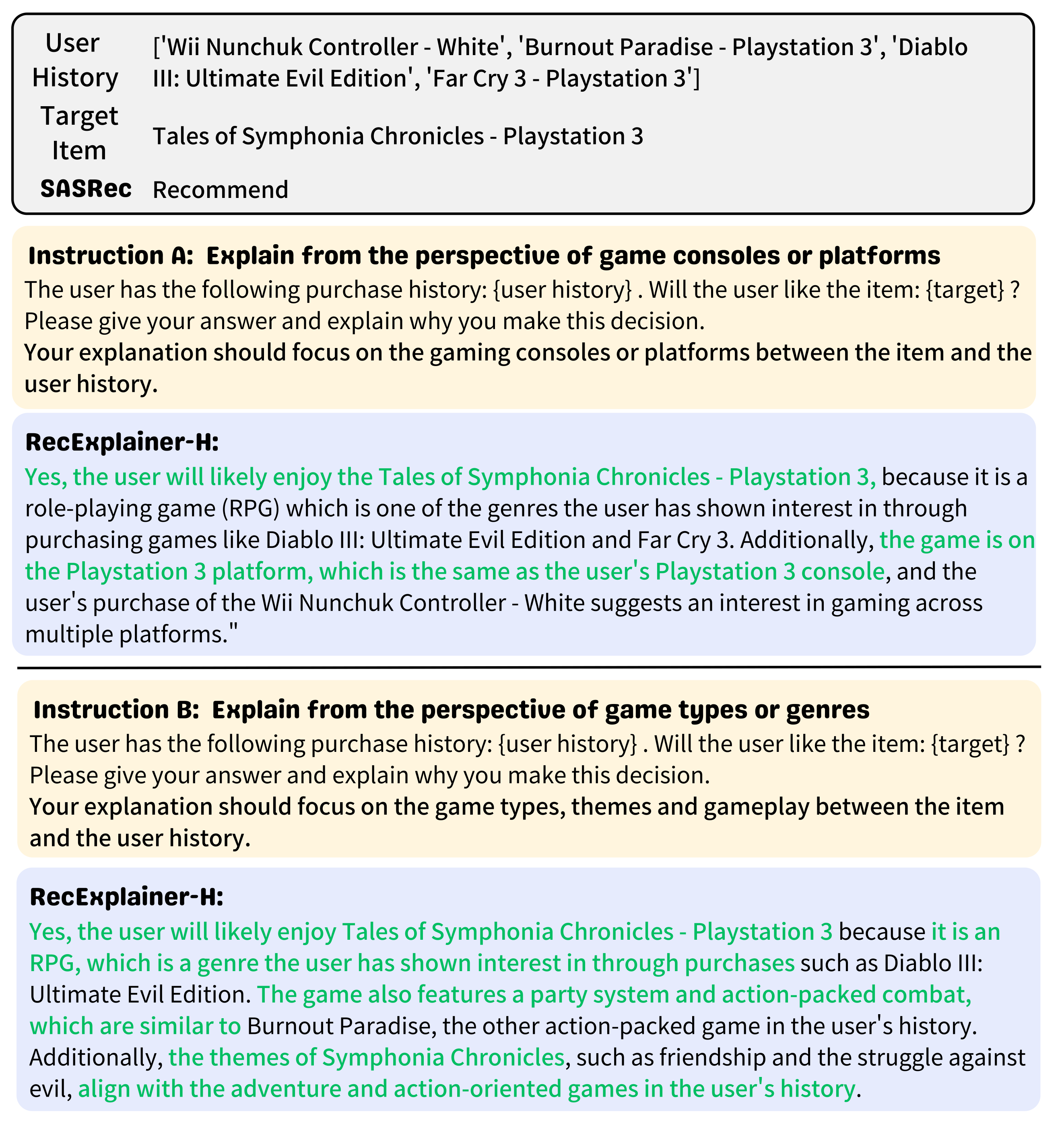}
     \vspace{-0.4cm}
    \caption{Case study for controllable explanation on Amazon Video Games dataset.}
    \label{fig:case_control}
\end{figure}

Benefit from the powerful instruction following capability and multi-step reasoning ability of LLM, our RecExplainer possesses interactive capabilities, enabling itself to understand user requirements and dynamically adjust the content of its explanations accordingly. Concretely, we instruct RecExplainer to predict and explain from two different view, i.e. the game platforms and game genres, the cases are shown in Figure~\ref{fig:case_control}. The model could generate consistent predictions with the target model in each case, and the two explanations indeed vary corresponding to the instructions, demonstrating the remained instruction following capability and the controllability of our RecExplainer.

\subsection{Ablation Study w.r.t Explanation}
\begin{figure}[t]
    \centering
    \includegraphics[width=1.0\columnwidth]{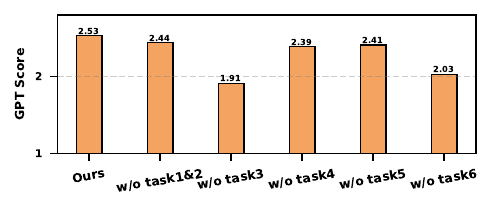}
    \vspace{-1cm}
    \caption{Ablation study on Amazon Video Games dataset.}
    \label{fig:ablation}
\end{figure}
In our method, we design multiple training tasks to help the LLM better understand target recommender models and domain-specific data. To explore the impact of each task on LLM's explanation generation ability, we remove each task to train a RecExplainer and evaluate the explanation quality using GPT-4. As shown in Figure \ref{fig:ablation}, each training task contributes to the final explanation quality. Specifically, task3 (interest classification) and task 6 (history reconstruction) have shown to have the most substantial impacts. On the one hand, compared to item embedding, information in user embedding is more severely compressed, and history reconstruction effectively facilitates learning this information. On the other hand, during explanation generation, the model needs to justify the prediction of the target model. Therefore, the classification task also has a significant impact on the model's explanation performance.

\section{Related Work}
\subsection{Model Explainability}
Deep neural models have demonstrated state-of-the-art (SOTA) performance in various machine learning tasks, and explaining the decision-making logic behind these black-box models has become a significant area of research. Existing literature in this field can be broadly divided into two categories~\cite{ying2019gnnexplainer}. The first category focuses on identifying the most salient parts of input features that contribute to the decision result of the target model, with the primary challenge being the effective implementation of score attribution.
\cite{erhan2009visualizing} present early work investigating the explainability of deep models by visualizing what a neuron computes in the input space, using several alternative methods such as sampling from a neuron and maximizing the activation of a neuron. Input space visualization has since been widely applied in vision tasks~\cite{zeiler2014visualizing,yosinski2015understanding} to facilitate human understanding of deep models. \cite{chen2018learning,xiting-xnlp} propose models that select salient input features by maximizing the mutual information between salient features and response variables. Additionally, other popular feature selection methods include neuron contribution difference~\cite{pmlr-v70-shrikumar17a}, integrated gradients~\cite{pmlr-v70-sundararajan17a}, influence functions~\cite{pmlr-v70-koh17a}, and Shapley value estimation~\cite{vstrumbelj2014explaining,datta2016algorithmic,lundberg2017unified}.

The second category of methods involves training a surrogate model to explain the target model. Surrogate models should maintain high fidelity to the target model's predictive patterns while also possessing explainable properties. Consequently, linear models and decision trees are the most commonly used surrogate models. \cite{schmitz1999ann,zilke2016deepred,lakkaraju2017interpretable} develop methods for generating a small number of compact decision rules as surrogates to explain the target model. ~\cite{ribeiro2016should} proposes a general model explanation framework that allows for the flexible selection of various surrogate model forms, such as linear models or decision trees.
However, surrogate models need to be structurally simple for explaining complex models, which creates a dilemma with model fidelity. Recently, LLMs have demonstrated strong versatility in terms of predictive accuracy and reasoning/explanation ability, providing an opportunity to overcome this dilemma. As a novel advancement in the first category of related works, \cite{bills2023language} utilizes GPT-4 to automatically generate explanations for neurons based on the attributed input features. To the best of our knowledge, this paper is the first to discuss leveraging LLMs for the second category of methods.

\subsection{LLMs and Multimodality}
In recent years, the Transformer architecture~\cite{vaswani2017attention} has become a fundamental building block for language models. Researchers have discovered that by pretraining large Transformer-based language models on extensive open data sources, these models exhibit enhanced capabilities in knowledge accumulation, multi-task learning, and few-shot or even zero-shot predictions~\cite{kenton2019bert,radford2018improving,radford2019language,brown2020language,dong2019unified}. Remarkably, when the model size reaches a certain scale (e.g., 170 billion parameters), language models exhibit emergent abilities \cite{wei2022emergent}, which are unforeseen phenomena, such as instruction following, reasoning, and problem-solving skills, indicating a preliminary step towards AGI. Consequently, researchers have begun to investigate whether emergent abilities can also be present in smaller-scale models (e.g., 7B models) if trained effectively. This inquiry has led to the development of several popular models, including OPT~\cite{zhang2022opt}, Llama~\cite{touvron2023llama}, Vicuna~\cite{chiang2023vicuna}, WizardLM~\cite{xu2023wizardlm}, and phi-1.5~\cite{li2023textbooks}. 

The intelligence of large models is not solely confined to text; other data modalities, such as images \cite{DBLP:conf/iclr/DosovitskiyB0WZ21,he2022masked} and audio \cite{radford2023robust}, can also benefit from large-scale pretraining. Multimodal models \cite{li2019visualbert,radford2021learning,DBLP:conf/cvpr/WangBDBPLAMSSW23,wang2022vlmixer} seek to break boundaries and bridge different modalities, as humans can simultaneously comprehend multimodal knowledge and perform complex tasks using all available information. Fundamentally, multimodal models align the models' representations across different modalities in the form of latent embeddings \cite{DBLP:conf/iclr/Wang0000YGW23,huang2023language}, enabling them to not only perceive each individual modality but also the interactions across modalities. With this perspective in mind, this paper considers the embeddings of recommender models as a new data modality and proposes a novel model explainer based on aligning LLMs.

\section{Conclusion}
In this paper, we investigate the potential of employing large language models (LLMs) as surrogate models to enhance the explainability of recommender systems. LLMs, known for generating high-quality, human-readable explanations, offer a promising solution to the traditional dilemma of necessitating simple models for self-explainability, thus paving the way for more advanced and transparent AI systems.  We introduce three innovative alignment approaches — behavior alignment, intention alignment, and hybrid alignment — to facilitate effective model alignment. Each of these approaches offers unique advantages in terms of explainability and fidelity. Through rigorous evaluation on three publicly available datasets, we demonstrate the effectiveness of our proposed alignment approaches in both comprehension and explanation. Empirical evidence highlights the potential of LLMs as a new type of surrogate model for explainable recommender systems. As an initial attempt, our research contributes to the ongoing efforts in explainable AI, paving the way for future work on leveraging LLMs for a wide range of explainability applications in complex systems.

\begin{acks}
The work was supported by grants from the National Key R\&D Program of China (No. 2021ZD0111801) and the National Natural Science Foundation of China (No. 62022077).
\end{acks}

\bibliographystyle{ACM-Reference-Format}
\balance
\bibliography{myref}

\appendix
\renewcommand\thefigure{\Alph{section}\arabic{figure}} 
\setcounter{figure}{0} 
\renewcommand\thetable{\Alph{section}\arabic{table}} 
\setcounter{table}{0}

\section{Details for human and GPT-4 annotations}\label{appendix:annotation}
\begin{figure*}
\centering
\small
\begin{mdframed}[backgroundcolor=gray!20]
Please act as an impartial judge and evaluate the AI assistant's recommendation decision as well as decision explanation based on the user's purchase history, target item, and ground truth label. Assign a score according to the following four levels:\\ \\
RATING-0: Incorrect classification - The assistant fails to generate a correct recommendation decision.\\ \\
RATING-1: Correct classification, insufficient explanation - The assistant correctly makes the recommendation decision but provides no, few, or irrelevant explanations, or provides explanations with hallucination, some of which do not conform to the actual situation.\\ \\
RATING-2: Correct classification, acceptable explanation - The assistant correctly makes the recommendation decision and provides an explanation that is logically consistent and aligns with the user's history and target item. But the explanation still has minor imperfections such as lack of persuasiveness or informativeness.\\ \\
RATING-3: Correct classification, satisfying explanation - The assistant correctly makes the recommendation decision and provides a satisfactory explanation, including a summary of the user's historical behavior patterns and characteristics, as well as a thorough analysis of the consistency or inconsistency between user preferences and the target item.\\ \\
Please give your score in the form of <br>RATING</br>, for example, if the rating is 1, output <br>RATING-1</br>. Do not allow the length of the explanation to influence your evaluation. Be as objective as possible.\\ \\
Known information: User history: \{USER HISTORY\}, Target item: \{ITEM\}, Label: \{YES/NO\}. Assistant's output: \{EXPLANATIONS\}
\end{mdframed}
\vspace{-0.2cm}
\caption{Prompt for the evaluation criteria.}
\label{fig: criteria}
\end{figure*}

\subsection{Criteria for GPT-4 and Human Experts}\label{appendix:criteria}

We ask human experts and GPT-4 to score explanations generated by all the LLMs with the same criteria. The prompt is shown in Figure \ref{fig: criteria}.

\subsection{Human Evaluation Setup}
We ask three experts, all of whom are master students majoring in recommender systems and are not involved in the co-author list, to evaluate the generated results of all LLMs.
These three experts coordinate the standards of the 4-level rating system before starting annotations and then each of them rates all the instances independently. During the evaluation process, they are presented with the target label, user history, target item, and model responses. Model responses are listed in random order, with all the model information anonymized, ensuring that the experts are unaware of the specific LLM responsible for generating each text.

\subsection{Human and GPT-4 Evaluation Agreement}
We have also included calculations for both inter-human agreement and gpt4-human agreement. When calculating inter-human agreement, we conduct pairwise comparisons among the three human annotators and compute the average metric. For the gpt4-human agreement calculation, we separately compute the metrics for the three gpt4-human pairs and then average them.

We first report Cohen's $\kappa$, which is commonly used to measure inter-rater agreement for categorical items. When calculating this, we treat the 4-level rating (1-4) as a categorical variable. The $\kappa$ for inter-human and gpt4-human is 0.366 and 0.316 respectively, which both show a moderate agreement.

We also compute the Spearman correlation coefficient $\rho$ between the ratings of our two evaluators (human or gpt4) by treating the rating as an ordinal variable (4>3>2>1). The coefficient for inter-human and gpt4-human is 0.563 and 0.714 respectively, which both indicate a high correlation between the two evaluators.

\section{Details about Dataset generation}
\label{appendix:data details}
\begin{table}[h]
  \caption{Statistics of the datasets.}
  \label{tab:datasets}
  \begin{tabular}{lrrrr}
    \toprule
    \textbf{Dataset} & \textbf{\#Users} & \textbf{\#Items} &\textbf{\#Inters} &\textbf{Sparsity} \\
    \midrule
    \textbf{Games} & 3,901& 1,864 & 31,672 &99.564\%\\
    \textbf{Movies}& 3,194 & 1,170 & 28,105 & 99.248\%\\
    \textbf{Steam}& 2,493 & 986 & 27,498 & 98.881\%\\
    \bottomrule
  \end{tabular}
\end{table}

\begin{table}[h]
  \centering
  \caption{Dataset details for each task.}
  \label{tab:dataset details}
  \setlength{\tabcolsep}{4.5pt} 
  \fontsize{7.5pt}{9.5pt}\selectfont
  \begin{tabular}{@{}l|lllll|lllll|lllll@{}}
    \toprule
    \multicolumn{1}{l|}{Dataset} & \multicolumn{2}{c|}{Video Games} & \multicolumn{2}{c|}{Movies and TV} & \multicolumn{2}{c}{Steam} \\ \midrule
    \multicolumn{1}{l|}{Split} & \multicolumn{1}{l}{\# Train} & \multicolumn{1}{l|}{\# Test} & \multicolumn{1}{l}{\# Train} & \multicolumn{1}{l|}{\# Test} & \multicolumn{1}{l}{\# Train} & \multicolumn{1}{l}{\# Test} & \\ \midrule
    \multicolumn{1}{l|}{Task 1} & \multicolumn{1}{c}{23,870} & \multicolumn{1}{c|}{3,901} & \multicolumn{1}{c}{21,717} & \multicolumn{1}{c|}{3,194} & \multicolumn{1}{c}{22,512} & \multicolumn{1}{c}{2,493} \\
    \multicolumn{1}{l|}{Task 2} & \multicolumn{1}{c}{23,870} & \multicolumn{1}{c|}{3,901} & \multicolumn{1}{c}{21,717} & \multicolumn{1}{c|}{3,194} & \multicolumn{1}{c}{22,512} & \multicolumn{1}{c}{2,493} \\
    \multicolumn{1}{l|}{Task 3} & \multicolumn{1}{c}{7,802} & \multicolumn{1}{c|}{2,294} & \multicolumn{1}{c}{6,388} & \multicolumn{1}{c|}{1,888} & \multicolumn{1}{c}{4,986} & \multicolumn{1}{c}{1,456} \\
    \multicolumn{1}{l|}{Task 4} & \multicolumn{1}{c}{9,177} & \multicolumn{1}{c|}{0} & \multicolumn{1}{c}{4,363} & \multicolumn{1}{c|}{0} & \multicolumn{1}{c}{3,856} & \multicolumn{1}{c}{0} \\
    \multicolumn{1}{l|}{Task 5} & \multicolumn{1}{c}{10,000} & \multicolumn{1}{c|}{1,000} & \multicolumn{1}{c}{10,000} & \multicolumn{1}{c|}{1,000} & \multicolumn{1}{c}{10,000} & \multicolumn{1}{c}{1,000} \\
    \multicolumn{1}{l|}{Task 6} & \multicolumn{1}{c}{27,771} & \multicolumn{1}{c|}{3,901} & \multicolumn{1}{c}{24,911} & \multicolumn{1}{c|}{3,194} & \multicolumn{1}{c}{25,005} & \multicolumn{1}{c}{2,493} \\ \midrule
    \multicolumn{1}{l|}{Total} & \multicolumn{1}{c}{102,490} & \multicolumn{1}{c|}{14,997} & \multicolumn{1}{c}{89,096} & \multicolumn{1}{c|}{12,470} & \multicolumn{1}{c}{88,871} & \multicolumn{1}{c}{9,935} \\
  \bottomrule
\end{tabular}
\end{table}

\subsection{Templates for Data Generation}\label{appendix:data generation}
To better align the LLM to the target recommendation model, we design several training tasks for the LLM to understand the predictive behaviors of the target model and the domain-specific data. Following \cite{chiang2023vicuna}, All tasks are formed into USER-ASSISTANT format. We list all the templates used in our datasets in Figure \ref{fig: prompt for data}.

\subsection{Statistics of the Datasets}\label{appendix:data stat}
Considering that each dataset has millions of interaction records, we reduce their sizes to avoid unacceptable training costs. Concretely, we filter each dataset by first selecting items with top frequency, and retaining users' interaction history on this item set. Finally, we randomly select a subset of users as our dataset. Following prior work \cite{DBLP:journals/corr/abs-2305-07001}, we also apply a 5-core filter, further removing users and items with fewer than five interactions from the dataset. We show the overall data statistics and statistics for each task in Table ~\ref{tab:datasets} and Table~\ref{tab:dataset details}, respectively.

\begin{figure*}[t]
    \centering
    \small
    \begin{mdframed}[backgroundcolor=gray!20]
        \textbf{Task Name}: Next Item Retrieval \\
        \textbf{Assistant Response}: \textit{"\{ITEM TITLE\}"} \\
        \textbf{User Question}: \\
        1. \textit{"Given the user purchase history: \{USER HISTORY\} , generate the next most likely clicked item title."}

        2. \textit{"What is the next most likely clicked item title for the purchase history: \{USER HISTORY\} ?"}
        
        3. \textit{"Predict the item that the user with this history: \{USER HISTORY\} might like next."}
        
        4. \textit{"Considering the purchasing history: \{USER HISTORY\} , what will be the next item the user click on?"}
        
        5. \textit{"Based on the buying history \{USER HISTORY\} , what item is the user likely to click on next?"}
        
        6. \textit{"With the given purchase records \{USER HISTORY\} , can you determine the next item the user will click?"}
        
        7. \textit{"What item is expected to be clicked next by a user who has this purchase history: \{USER HISTORY\} ?"}
        
        8. \textit{"Generate the next probable clicked item for a user with the purchase history: \{USER HISTORY\} ."}
        
        9. \textit{"For a user with the following purchase background: \{USER HISTORY\} , which item will he most likely click next?"}
        \\

       \textbf{Task Name}: Item Ranking \\
        \textbf{Assistant Response}: \textit{"\{SORTED ITEM TITLES\}"} \\
        \textbf{User Question}: \\
        1. \textit{"Given the user history: \{USER HISTORY\} and next items to be ranked: \{ITEMS\} , generate the sorted item titles from the user's favorite to least favorite."}

        2. \textit{"Considering user: \{USER HISTORY\} and some items he might like next: \{ITEMS\} , provide a ranking list of them according to the user preference."}
        
        3. \textit{"Please rank the following items: \{ITEMS\} from what the user likes to dislikes. Here is the user history: \{USER HISTORY\} ."}
        
        4. \textit{"For user with purchase history: \{USER HISTORY\} , please arrange these items in order of preference: \{ITEMS\} ."}
        
        5. \textit{"Taking into account user's history: \{USER HISTORY\} , create a list of the items: \{ITEMS\} ranked by the user's interests."}
        
        6. \textit{"With the user's purchase history given: \{USER HISTORY\} , sort the items: \{ITEMS\} based on the user's taste from best to worst."}
        
        7. \textit{"Based on the purchase history: \{USER HISTORY\} , please provide a ranking of the following items: \{ITEMS\} according to the user's preferences."}
        
        8. \textit{"Given user's past history: \{USER HISTORY\} , rank these items: \{ITEMS\} from most to least appealing."}
        
        9. \textit{"Using the provided user purchase history: \{USER HISTORY\} , generate a ranked list of items: \{ITEMS\} in accordance with the user's likes and dislikes."}
        \\

        \textbf{Task Name}: Interest classification \\
        \textbf{Assistant Response}: \textit{"\{YES/NO\}"} \\
        \textbf{User Question}: \\
        1. \textit{"The user has the following purchase history: \{USER HISTORY\} . Will the user like the item: \{ITEM\} ?"} 

        2. \textit{"Considering user: \{USER HISTORY\} and item: \{ITEM\} , will the user like the item?"} 

        3. \textit{"Here is the user history: \{USER HISTORY\} . Do you think the user will prefer the item: \{ITEM\} ?"} 

        4. \textit{"User's purchase records are: \{USER HISTORY\} . Can you tell if the user will enjoy item: \{ITEM\} ?"} 

        5. \textit{"Given the purchase background of the user: \{USER HISTORY\} , would the user appreciate the item: \{ITEM\} ?"} 

        6. \textit{"The buyer has this purchase history: \{USER HISTORY\} . Would the user be interested in the product: \{ITEM\} ?"} 

        7. \textit{"With the following purchasing history for the user: \{USER HISTORY\} , can we predict if the user will like item: \{ITEM\} ?"} 

        8. \textit{"Here's the customer's buying log: \{USER HISTORY\} . Would you say the user might favor the item: \{ITEM\} ?"} 
        \\
        
        \textbf{Task Name}: Item discrimination\\
        \textbf{Assistant Response}: \textit{"\{TITLE/DESCRIPTION/TAGS/SIMILAR ITEM TITLE/BRAND/FEATURE\}"} \\
        \textbf{User Question}: \\
        1. \textit{"What is the \{TITLE/DESCRIPTION/TAGS/SIMILAR ITEM TITLE/BRAND/FEATURE\} of the item: \{ITEM\} ?"}

        2. \textit{Given the item: \{ITEM\} , generate its \{TITLE/DESCRIPTION/TAGS/SIMILAR ITEM TITLE/BRAND/FEATURE\}."}

        3. \textit{For the item: \{ITEM\} , what is its \{TITLE/DESCRIPTION/TAGS/SIMILAR ITEM TITLE/BRAND/FEATURE\}?"}

        4. \textit{"Can you tell me the \{TITLE/DESCRIPTION/TAGS/SIMILAR ITEM TITLE/BRAND/FEATURE\} of the item: \{ITEM\} ?"}

        5. \textit{"Please generate the \{TITLE/DESCRIPTION/TAGS/SIMILAR ITEM TITLE/BRAND/FEATURE\} of the item: \{ITEM\} ."}

        6. \textit{\{TITLE/DESCRIPTION/TAGS/SIMILAR ITEM TITLE/BRAND/FEATURE\} of the item: \{ITEM\} ?"}

        7. \textit{"Item: \{ITEM\} , what is its \{TITLE/DESCRIPTION/TAGS/SIMILAR ITEM TITLE/BRAND/FEATURE\}?"}

        8. \textit{"Could you generate the \{TITLE/DESCRIPTION/TAGS/SIMILAR ITEM TITLE/BRAND/FEATURE\} for the item: \{ITEM\} ?"}
        \\

        \textbf{Task Name}: History reconstruction\\
        \textbf{Assistant Response}: \textit{"\{USER HISTORY TITLES\}"} \\
        \textbf{User Question}: \\
        1. \textit{What are the history titles of the user: \{USER HISTORY\} ?"}

        2. \textit{Given the user purchase history: \{USER HISTORY\} , generate the history titles."}

        3. \textit{Generate the titles of the user history: \{USER HISTORY\} ."}

        4. \textit{Show me the history titles for the user: \{USER HISTORY\} ."}

        5. \textit{Can you list the titles in the purchase history of the user: \{USER HISTORY\} ?"}

        6. \textit{Please generate the titles from the user's purchase history: \{USER HISTORY\} ."}

        7. \textit{List the titles in the purchase history for user: \{USER HISTORY\} ."}

        8. \textit{What titles can be found in user's purchase history: \{USER HISTORY\} ?"}



    \end{mdframed}
    \vspace{-0.3cm}
    \caption{Prompt templates for data generation.}
    \label{fig: prompt for data}
\end{figure*}

\end{document}